\documentclass[twocolumn,secnumarabic,amssymb, nobibnotes, aps,graphicx]{revtex4-1}
\usepackage{amssymb}
\usepackage{bm}
\usepackage{graphicx}
\usepackage{color}
\evensidemargin \oddsidemargin
\def\0{{\bf 0}}

\def \U {{\cal U}}

\newcommand{\bi}{{\bf i}}
\newcommand{\bjj}{{\bf j}}
\newcommand{\bk}{{\bf k}}

\newcommand{\bb}{\mbox{\boldmath $\beta$}}
\newcommand{\Be}{\mbox{\boldmath $\epsilon$}}
\newcommand{\Bep}{\mbox{\boldmath $\epsilon$}^{\scriptscriptstyle \perp}}

\newcommand{\Bp}{{\bf p}}
\newcommand{\Bpp}{{\bf p}^{\scriptscriptstyle \perp}}

\newcommand{\C}{{\cal C}}
\newcommand{\D}{{\cal D}}

\newcommand{\bx}{{\bf x}}
\newcommand{\bxp}{{\bf x}^{\scriptscriptstyle \perp}}

\newcommand{\bX}{{\bf X}}
\newcommand{\bXp}{{\bf X}^{\scriptscriptstyle \perp}}

\newcommand{\bu}{{\bf u}}
\newcommand{\bup}{{\bf u}^{\scriptscriptstyle \perp}}

\newcommand{\bv}{{\bf v}}

\newcommand{\bj}{{\bf j}}
\newcommand{\bjp}{{\bm j}^{\scriptscriptstyle \perp}}
\newcommand{\bE}{{\bf E}}
\newcommand{\bB}{{\bf B}}
\newcommand{\bEp}{{\bf E}^{\scriptscriptstyle \perp}}

\newcommand{\bA}{{\bf A}}
\newcommand{\bAp}{{\bf A}\!^{\scriptscriptstyle \perp}}

\newcommand{\Bap}{{\bm \alpha}^{\scriptscriptstyle \perp}}

\def \E {{\cal E}}

\def \sm {s_{{\scriptscriptstyle -}}}
\def \sp {s_{{\scriptscriptstyle +}}}
\def \spm {s_{{\scriptscriptstyle \pm}}}

\newcommand{\be}{\begin{equation}}
\newcommand{\ee}{\end{equation}}
\newcommand{\bea}{\begin{eqnarray}}
\newcommand{\eea}{\end{eqnarray}}
\newcommand{\ba}{\begin{array}}
\newcommand{\ea}{\end{array}}

\begin{document}

\title{On cold diluted plasmas hit by short laser pulses}

\author{Gaetano Fiore$^{a,b}$, \ \   Paolo Catelan$^{c,d}$ \\
$^{a}$ Dip. di Matematica e Applicazioni, Universit\`a ``Federico II'', Complesso MSA, Via Cintia, 80126 Napoli, Italy\\
$^{b}$       INFN, Sezione di Napoli, Complesso  Universitario  M. S. Angelo,  Via Cintia, 80126 Napoli, Italy\\
$^{c}$ 
Centro de Energ\'ias Alternativas y Ambiente, 
Escuela Superior Polit\'ecnica del Chimborazo, Riobamba, Ecuador\\
$^{d}$ Dip. di Matematica ed Informatica, 
Universit\'a della Calabria, Arcavacata, Rende, Italy
}

\begin{abstract}
{\bf Abstract}: \ Adapting a plane hydrodynamical model we briefly  
revisit the study of the impact  of a very short and intense laser pulse 
onto a diluted plasma, the formation of a plasma wave, its wave-breaking, the occurrence of the slingshot effect.

{\bf Keywords}: \
Laser-plasma interactions; electron acceleration; plasma wave; wave-breaking; slingshot effect
\end{abstract}

\maketitle

\section{Introduction}
\noindent

The laser-plasma interactions induced by ultra-intense or ultra-short laser pulses are at the base of 
the Laser Wake-Field  (LWF)  \cite{Tajima-Dawson1979} and other extremely compact acceleration mechanism
of charged particles with extremely important potential  applications  in medicine, industry, etc.
The equations ruling them are so  complex that recourse to numerical resolution based 
on e.g. particle-in-cell (PIC)  techniques is almost unavoidable. PIC or other codes in general involve huge and costly computations for each choice of the free parameters; exploring the parameter space blindly to pinpoint  the interesting regions is  prohibitive, if not accompanied by some analytical insight that can simplify
the work, at least in special cases or in a limited space-time region.
This applies in particular to the impact of a very intense and  short laser pulse (the {\it pump}) on a cold diluted plasma (or matter to be locally  ionized  into a plasma by the pulse itself). 
Here we briefly revisit it using an improved 1D Lagrangian model: with very little computational power we can
more easily determine conditions (on the initial density $\widetilde{n_0}$, the laser length $l$ and spot radius $R$) for,
and information on:
i) the formation  of a plasma wave (PW)  \cite{AkhPol56,BulKirSak89} and its breaking \cite{Daw59} - if any - at  density inhomogeneities  \cite{BulEtAl98,BraEtAl08,LiEtAl13} (this is important as a possible injection mechanism for the LWFA);
ii)  the slingshot effect, i.e. the backward expulsion of a bunch of high-energy electrons from the plasma surface  
\cite{FioFedDeA14,FioDeN16} shortly after the impact of the pulse.
More detailed arguments will appear in the longer paper \cite{Fio17}.

We assume the plasma is initially neutral, unmagnetized and at rest with zero densities
in the region  \ $z\!<\! 0$. We describe it as  a static background of ions 
and a fully relativistic collisionless fluid of electrons, with plasma and electric, magnetic fields 
$\bE,\bB$ fulfilling the Lorentz-Maxwell and continuity equations.  We check
{\it a posteriori} where and how long  such a hydrodynamical picture 
is self-consistent.  We assume
 $t\!=\!0$ initial conditions for  the electrons  Eulerian density $n_e$, velocity $\bv_e$
 of the type
\be 
\bv_e(0 ,\!\bx)\!=\!\0, \qquad n_e(0,\!\bx)\!=\!\widetilde{n_0}(z), \quad \widetilde{n_0}(z)\!=\!0
\:\:\mbox{if }\: z\!\le\! 0,
 \label{asyc}
\ee
where  $0\!<\!\widetilde{n_0}(z)\!\le\! n_b$ if $z\!>\!0$ and
for simplicity  $\widetilde{n_0}(z)\!=\!n_0$ if $z\!\ge\! z_s$, for some
$n_b\!\ge\!n_0\!>\!0$,  $z_s\!\ge\! 0$.
As we regard ions as immobile, the proton density will be $ n_p(t,\bx)\!=\!\widetilde{n_0}(z)$ for all $t$.
We assume that the pump can be schematized for $t\!<\!0$ as a free plane transverse wave  $\Be^{{\scriptscriptstyle\perp}}\!(ct\!-\!z)$ \   traveling  in the $z$-direction
($\perp$ means orthogonal to  $\vec{z}$, $c$ is the speed of light) 
multiplied by a `cutoff' function  $\chi\!_{{\scriptscriptstyle R}}(\rho)$,
\be
\bEp (t, \bx)=\Be^{{\scriptscriptstyle\perp}}\!(ct\!-\!z)\,
\chi\!_{{\scriptscriptstyle R}}(\rho),\quad  \bB^{{\scriptscriptstyle\perp}}=
{\hat{\bf z}}\!\times\!\bEp \qquad \mbox{if } t\le 0;               
      \label{pump}
\ee
more precisely,  $\chi\!_{{\scriptscriptstyle R}}(\rho)$ is 1 if $\rho\!\equiv\!\sqrt{\!x^2\!+\!y^2}\!\le\! R$ 
and rapidly goes to zero for $\rho\!>\! R$, \ 
while $\Be^{{\scriptscriptstyle\perp}}\!(\xi)\!=\!0$ unless 
$0\!<\!\xi\!<\!l$, where the effective pulse length $l$ fulfills\footnote{If 
the plasma is created locally by the impact  of the pulse itself on a gas  (e.g. helium) jet, then
 $[0,l]$ is the interval where the intensity is sufficient to ionize the gas.
If $\chi\!_{{\scriptscriptstyle R}}(\rho)\!\neq\!1$ then (\ref{pump}), as similar idealizations 
used in the literature, violates the Maxwell equations, but is justified for short time lapses, in which it differs little
from a solution.
}
\vskip-.3cm
\be
2l\lesssim c t_{{\scriptscriptstyle H}},                             \label{Lncond}
\ee
\vskip-.1cm \noindent
and $t_{{\scriptscriptstyle H}}$  is  the plasma period associated to the density
$n_b$ [see (\ref{period}) below]. $\Be^{{\scriptscriptstyle\perp}}\!(\xi)\!=\!0$ if  $\xi\!<\!0$ means that the pulse reaches the plasma at $t\!=\!0$. 
Condition (\ref{Lncond})  secures that the pulse is completely inside the bulk before any electron gets out of it
 and is fulfilled if $l$ or  $n_b$ are small enough ({\it a fortiori} the plasma is underdense); 
in particular, if $2l\!\le\! ct_{{\scriptscriptstyle H}}^{{\scriptscriptstyle nr}}$; \ $t_{{\scriptscriptstyle H}}^{{\scriptscriptstyle nr}}\!\equiv\!\sqrt{\pi m/n_b e^2}\!\le\! t_{{\scriptscriptstyle H}}$  is the non-relativistic limit of $t_{{\scriptscriptstyle H}}$ ($m,-e$ are the electron mass, charge). As we make no extra assumptions on the Fourier spectrum or the polarization of $\Be^{{\scriptscriptstyle\perp}}$, our method can be applied  to all kind of such travelling waves, ranging from almost monochromatic to so called ``impulses".

 In section \ref{Planewavessetup} we discuss  the motion of electrons 
when $R\!=\!\infty$ ($\chi\!_{{\scriptscriptstyle R}}\!\equiv\! 1$) 
using an improved  plane hydrodynamical model \cite{Fio14JPA,Fio17JPA}  (for shorter presentations see \cite{FioRev}) that allows to
 reduce the system of Lorentz-Maxwell and continuity 
partial differential equations (PDEs)  into ordinary differential equations (ODEs), more
precisely into a family of decoupled systems of non-autonomous Hamilton 
Equations in dim 1 in rational form. In the model    
we alternatively adopt the light-like coordinate \ $\xi=ct\!-\!z$ \ or time $t$ to parametrize the electron motion, the transverse  and the light-like components 
\ $\Bpp_e$, $p_e^0\!-\!cp_e^z\!\equiv\! mc^2 s_e$  (instead of the longitudinal one $p_e^z$) of the electron 4-momentum as unknowns, neglect pump depletion, control how long this 
is valid, how long the hydrodynamical picture holds, 
when and where it fails (by wave-breaking \cite{Daw59}). Then we test the model
by numerically solving the ODE's with $\widetilde{n_0}(Z)$ either step-like, or as in \cite{BraEtAl08}
(i.e. linearly growing in a first interval and decreasing in a second),
and find consistent results.
In section \ref{3D-effects} we use causality and heuristic arguments to qualitatively adapt these results
to  the ``real world" ($R\!<\!\infty$) and justify the above statements.

\section{Set-up and plane model}
\label{Planewavessetup}

\begin{figure}
\includegraphics[width=7.8cm]{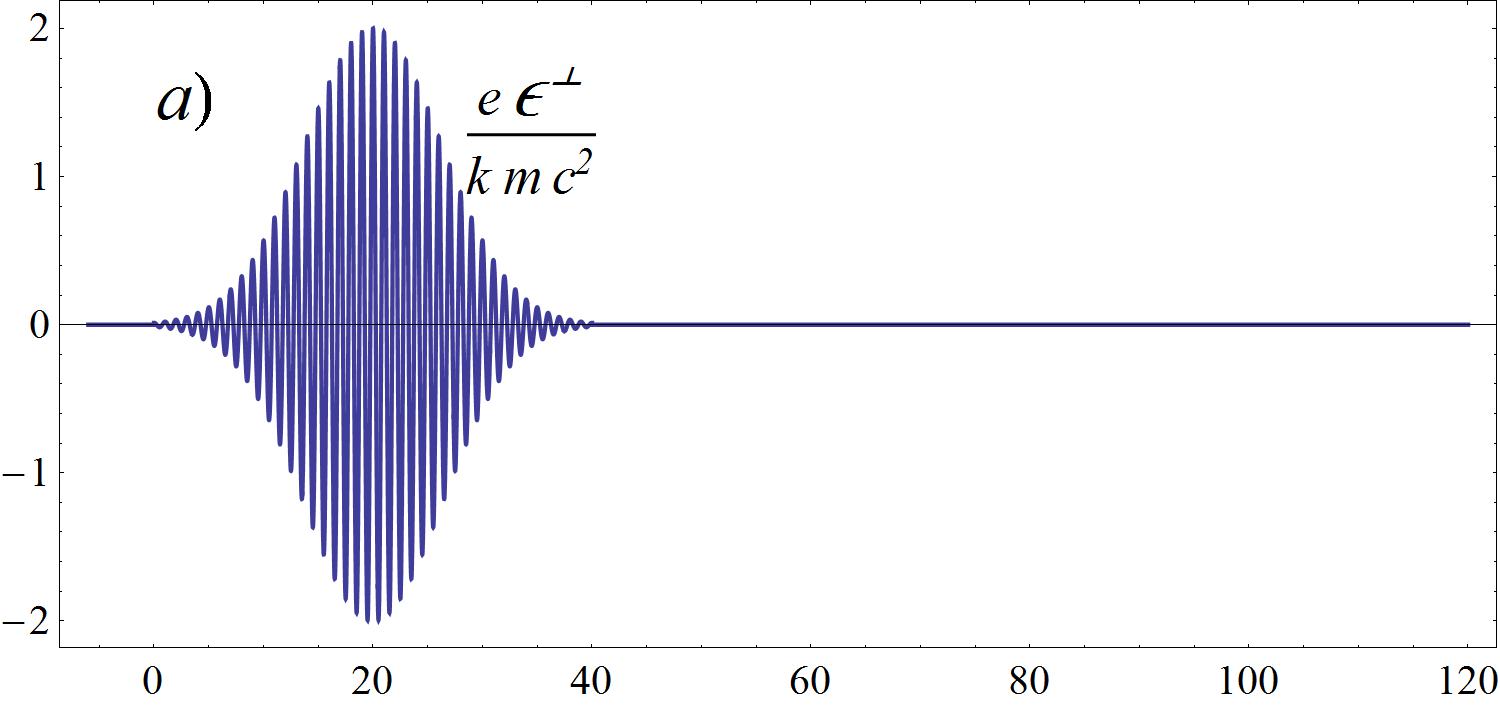}\\
\includegraphics[width=7.8cm]{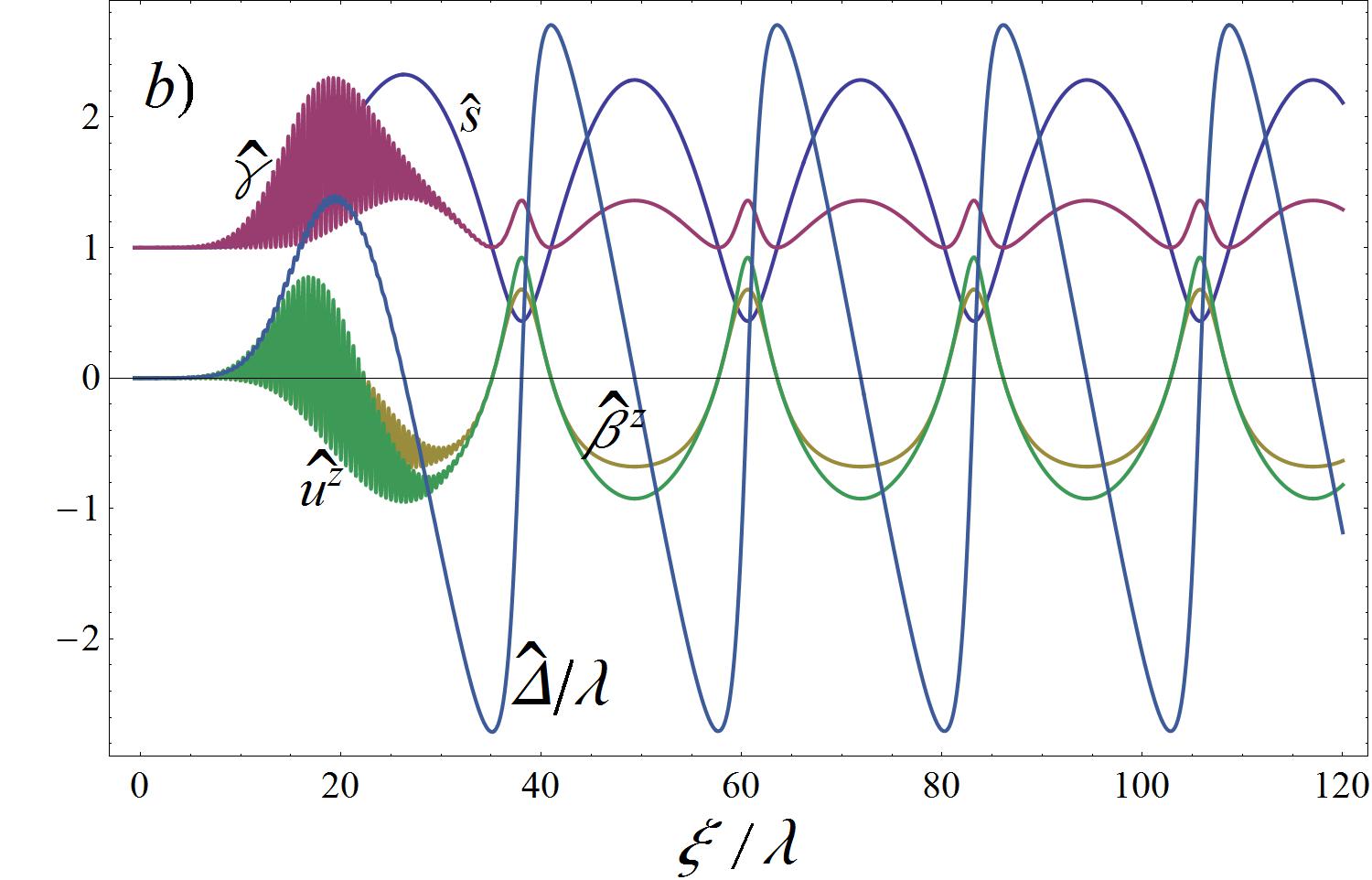} \\
\includegraphics[width=7.8cm]{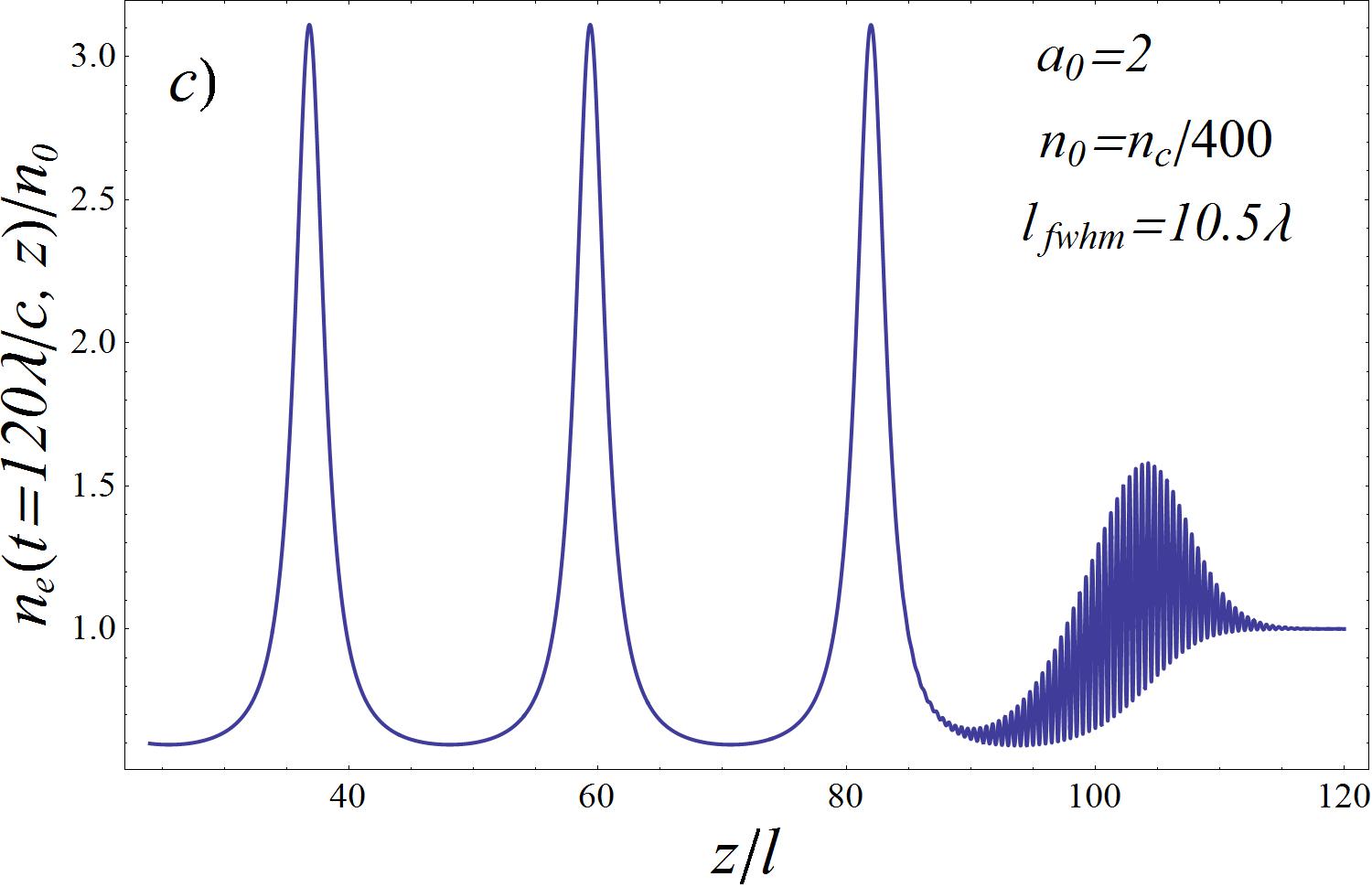}\\
\includegraphics[width=7.8cm]{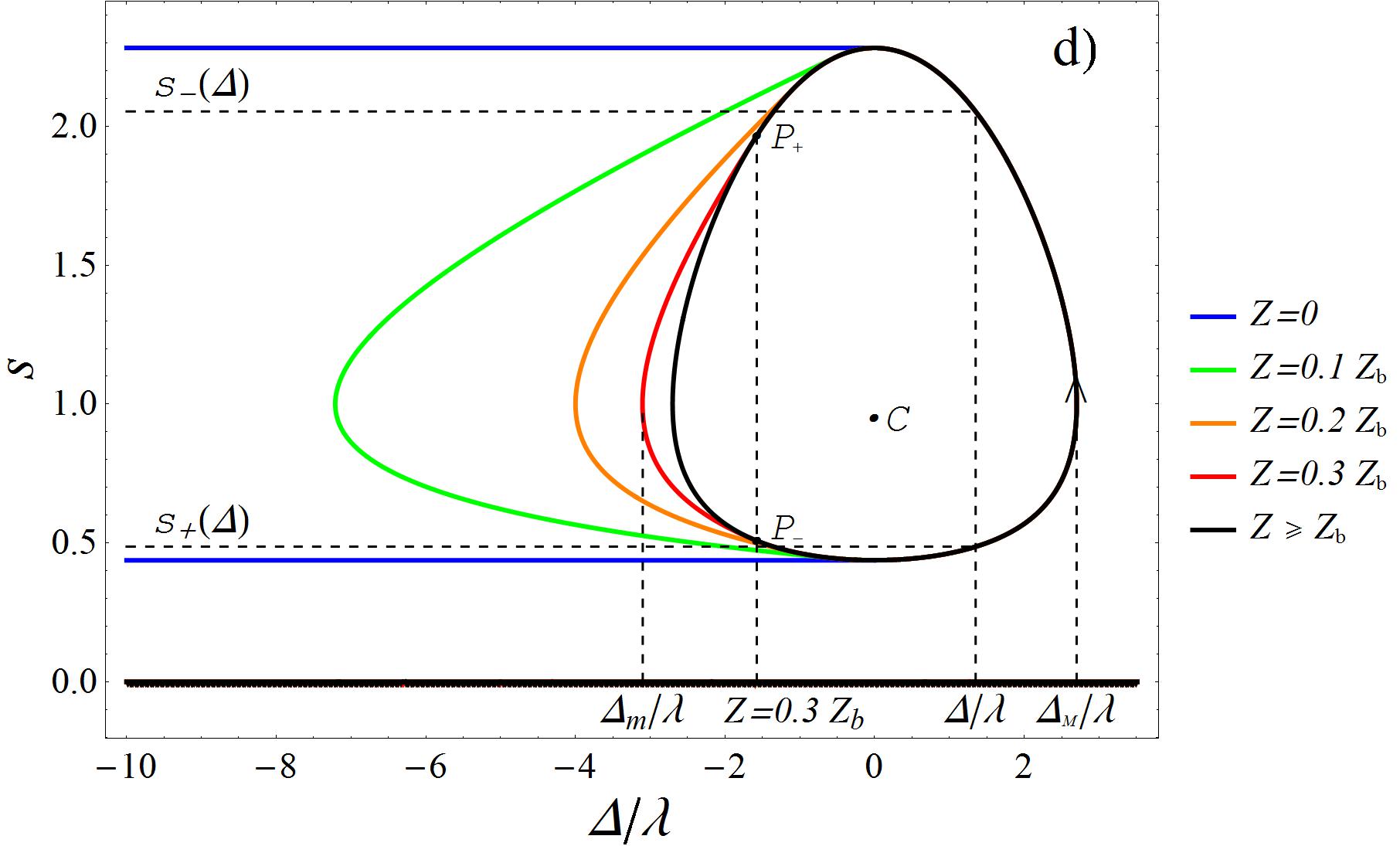}
\caption{
(Color online) 
\ a)  Normalized gaussian pump of length $l_{fwhm}\!=\!10.5\lambda$, linear polarization, 
peak  amplitude $a_0\!\equiv\!2\pi eE^{\scriptscriptstyle \perp}_{\scriptscriptstyle M}/mc^2k\!=\!2$
(leading to a peak  intensity $I\!=\!1.7\!\times\!10^{19}$W/cm$^2$ if $\lambda=0.8\mu$m). \ \ b) \ Corresponding  
solution of (\ref{e1}-\ref{heq2}) for $Z\!\ge\!Z_d$, with
$\widetilde{n_0}(Z)\!=\!n_0\theta(Z)$, $n_0\!\equiv\! n_{cr}/400$
  (whence $h\!=\!1.36$); as anticipated, $\hat s$ is indeed  insensitive to fast oscillations of $\Bep$. \
c) \ Corresponding normalized electron density inside the bulk as a function of $z$ at \ $ct=120\lambda$. \ \ 
d) \ Phase portraits for  the same $\widetilde{n_0}(Z)$, $h\!=\!1.36$, $\mu=1$. These values of 
$l_{fwhm}$, $a_0$, $n_0$ are picked from \cite{BraEtAl08}.
}
\label{graphsb}
\end{figure}

\begin{figure*}[ht]
\begin{center}
\includegraphics[width=13.3cm]{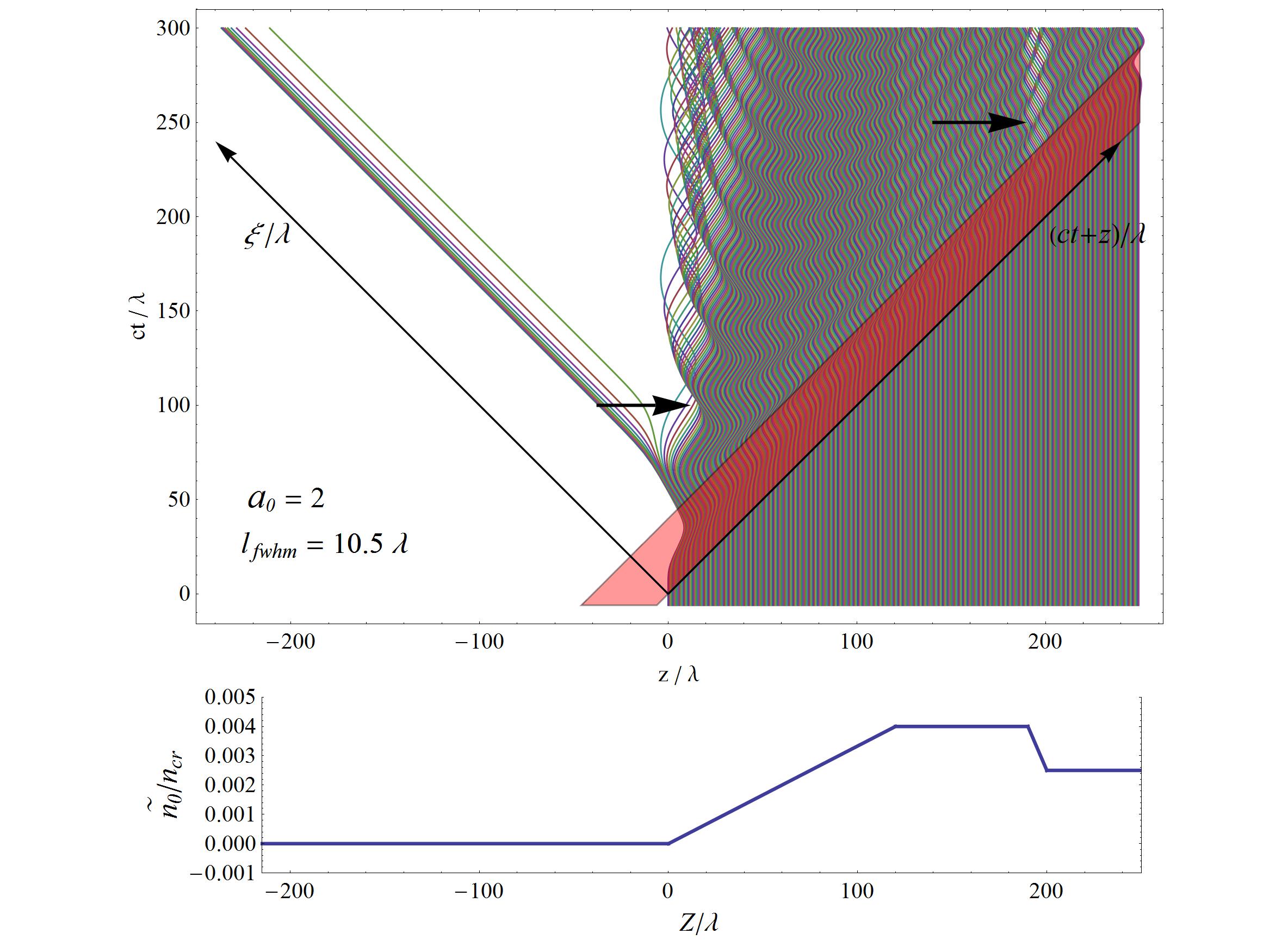}\hfill
\includegraphics[width=3.9cm]{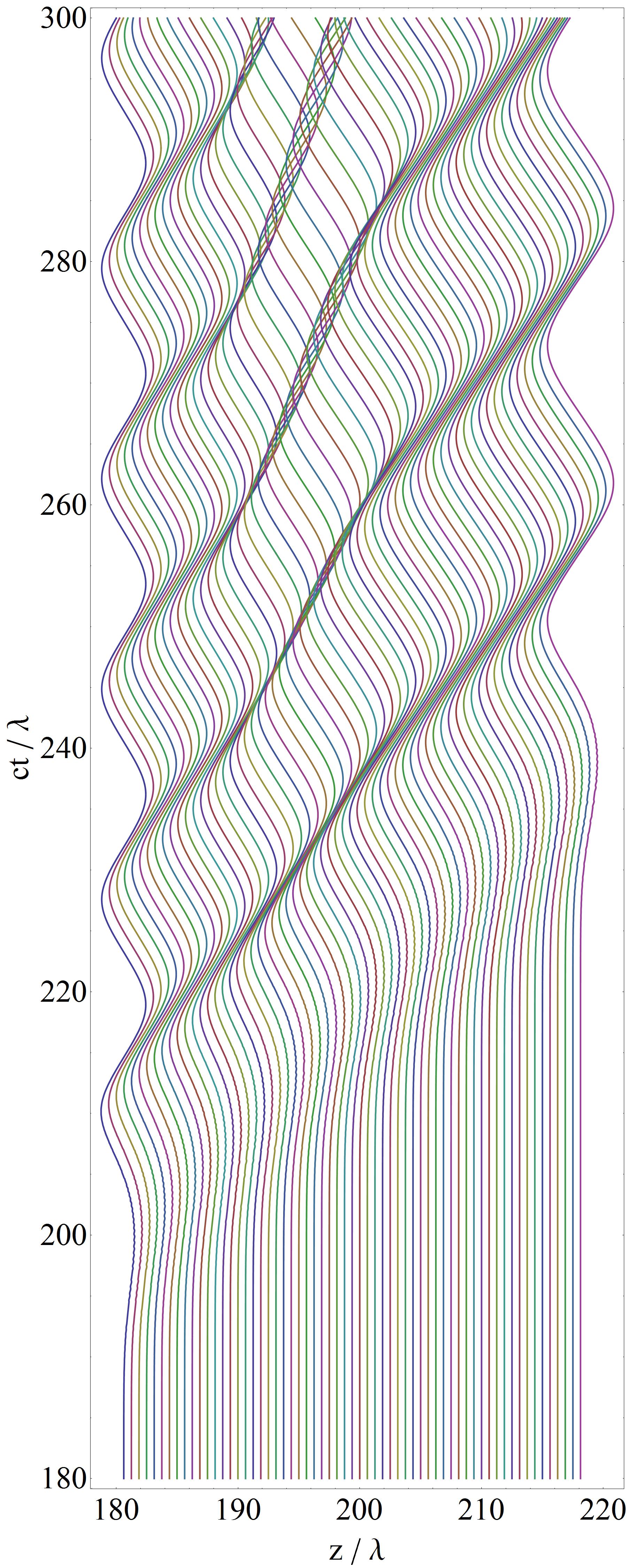}
\end{center}
\caption{The electron worldlines (WL) induced by the pulse on an initial density $\widetilde{n_0}(Z)$ as in section III.B of Ref. \cite{BraEtAl08}: WL of $Z\!\sim\!0$ electrons stray left away,
WL of other up-ramp electrons first intersect after about $5/4$ oscillations (left arrow
), WL of down-ramp electrons first intersect after about  $7/4$ oscillations 
(right arrow; 
see also the higher resolution plot at the right),
consistently with the results obtained in \cite{BraEtAl08} by a 2D PIC simulation with 
a gaussian $\chi\!_{{\scriptscriptstyle R}}(\rho)$ with  FWHM equal to $20\lambda$. 
The  support of $\Bep(ct\!-\!z)$  is pink (we have
considered $\Bep(\xi)\!=\!0$ outside $0\!<\!\xi\!<\!40\lambda$).
}
\label{Worldlinescrossings}       
\end{figure*}

The  equations of motion of an electron $e^-$
is non-autonomous and highly nonlinear in the unknowns the position \ $\bx(t)$ \ and momentum \ $\Bp(t)=mc\,\bu(t)$:
\bea
\ba{l}
\displaystyle\dot\Bp(t)=-e\bE[t,\bx(t)] - \frac{\Bp(t) \wedge e\bB[t,\bx(t)]  }{\sqrt{m^2c^2\!+\!\Bp^2(t)}} 
,\\[6pt]  
\displaystyle
\frac{\dot \bx(t)}c =\frac{\Bp(t) }{\sqrt{m^2c^2\!+\!\Bp^2(t)}} ,
\ea
\label{EOM}
\eea
We decompose $\bx\!=\!x\bi\!+\!y\bjj\!+\!z\bk\!=\bxp\!+\!z\bk$, etc, in the cartesian coordinates of the laboratory frame, and often use the dimensionless variables $\bb\!\equiv\!\bv/c\!=\!\dot\bx/c$,  
\  $\gamma\!\equiv\! 1/\sqrt{1\!-\!\bb^2}\!=\!\sqrt{1\!+\! \bu^2}$, \ 
the 4-velocity $u\!=\!(u^0\!,\bu)\!\equiv\!(\gamma,\gamma \bb)
$, i.e. the dimensionless version of the 4-momentum $p$. As by (\ref{EOM}b) $e^-$ cannot reach the speed of light, \ $\tilde \xi(t)\!=\!ct\!-\!z(t)$  grows strictly,
and we can make the change  $t\mapsto \xi\!=\!ct\!-\!z$ of independent parameter  along the worldline  
of $e^-$ (see fig. \ref{Worldlinescrossings}), so that the term $\Bep[ct-z(t)]$, where the {\it unknown} $z(t)$ is in the argument of the highly nonlinear and rapidly varying $\Bep$, becomes the {\it known} forcing term $\Bep(\xi)$. We denote as $\hat \bx(\xi)$ the position of $e^-$ as a function of $\xi$; it is determined  by $\hat \bx(\xi)=\bx(t)$. More generally we 
denote $\hat f(\xi, \hat \bx)\equiv f[(\xi\!+\!  \hat z)/c,  \hat \bx]$ for any given function $f(t,\bx)$,
 abbreviate $\dot f\!\equiv\! df/dt$, $\hat f'\!\equiv\! d\hat f/d\xi$ (total derivatives). 
It is convenient to make also the change of dependent (and unknown) variable $u^z\!\mapsto\! s$, where the {\it $s$-factor}  \cite{Fio17}
\be
s\equiv\gamma\!- u^z=u^-=\gamma(1-\beta^z)=\frac{\gamma}c \frac{d\tilde \xi}{dt}>0          \label{defs0}
\ee
is the light-like component of $u$, as well as the Doppler factor of $e^-$:
$\gamma,\bu,\bb$ are the {\it rational}  functions of $\bu^{{\scriptscriptstyle\perp}}, s$
\be
\gamma\!=\!\frac {1\!+\!\bu^{{\scriptscriptstyle\perp}}{}^2\!\!+\!s^2}{2s}, 
\qquad  u^z\!=\!\frac {1\!+\!\bu^{{\scriptscriptstyle\perp}}{}^2\!\!-\!s^2}{2s}, 
 \qquad  \bb\!=\! \frac{\bu}{\gamma}                                    \label{u_es_e}
\ee
(these relations hold also with the carets); so, replacing $d/dt\mapsto(c s/ \gamma)d/d\xi$ and putting carets  on all variables (\ref{EOM}) becomes {\it rational} in the unknowns $\hat\bu^{{\scriptscriptstyle\perp}},\hat s$, in particular  (\ref{EOM}b) becomes $\hat\bx'=\hat\bu/\hat s$. Moreover, $\hat s$ is practically insensitive to fast oscillations
of $\Bep(\xi)$ (as e.g. fig. \ref{graphsb}b illustrates).
Passing to the plasma, we denote as \ $\bx_e(t,\bX)$  \ the position at time $t$
of the electrons' fluid element $d^3X$ initially located at $\bX\!\equiv\!(X,Y,Z)$, as $\hat \bx_e(\xi,\bX)$ the position   of $d^3X$  as a function of $\xi$.
 For brevity   we refer to the electrons initially contained: in $d^3X$, as the  `$\bX$ electrons';  in a 
region $\Omega$, as the `$\Omega$ electrons'; in the layer between $Z,Z\!+\!dZ$,
as  the `$Z$ electrons'.
The map $\bx_e(t, \cdot):\bX\mapsto \bx$ must be one-to-one for every $t$;
equivalently,   $\hat\bx_e(\xi, \cdot):\bX\mapsto \bx$ must be one-to-one  for every $\xi$.
Clearly, 
\be
\bX_e(t,  \bx)=\hat \bX_e(ct\!-\!z,  \bx). \label{clear}
\ee

In this section we set $\chi\!_{{\scriptscriptstyle R}}\!\equiv\! 1$  in (\ref{pump}),
so  that all initial data are independent of transverse coordinates. Hence, 
also the Eulerian fields 
solving the equations will depend only on \  $z,t$, thus justifying the Ansatz
$A(t,z)$ for the EM potential  $A\!=\!(A^0\!,\!\bA)$. Then the initial condition (\ref{pump}) 
 implies that for all $t$
$\bAp$ is a physical observable and
\be
\bA\!^{{\scriptscriptstyle\perp}}\!(t\!,\!z)\!=\!- c\!\!\!\int\limits^{t}_{ -\infty }\!\!\!  dt'  \bEp\!(t{}'\!,\!z),
\quad c\bEp\!\!=\!-\partial_t\bAp,\quad \bB\!=\!\bk\!\wedge\!\partial_z\bAp\!.   \label{Ap}
\ee
Similarly, the displacement  \ $\Delta \bx_e\equiv \bx_e(t,\!\bX\!) -\!\bX$ will 
actually depend only on $t,Z$ [and $\Delta \hat\bx_e\equiv \hat\bx_e(\xi,\bX)\!-\!\bX$ only on $\xi,Z$]
and by causality vanishes  if \  $ct\!\le\! Z$.
The Eulerian electrons' momentum $\Bp_e(t,z)$ obeys equation
(\ref{EOM}), where one has to replace $\bx(t)\mapsto\bx_e(t,\bX)$,
 $\dot \Bp\mapsto d\Bp_e/dt\equiv$ {\it total} derivative; as known, by (\ref{Ap}) 
the transverse part of (\ref{EOM}a)  becomes
$\frac{d\Bp^{{\scriptscriptstyle\perp}}_e}{dt}=\frac ec \frac{d\bAp}{dt}$, 
which due to $\Bp^{{\scriptscriptstyle\perp}}_e(0,\bx)=\0$ implies
\be
\Bp^{{\scriptscriptstyle\perp}}_e=\frac ec \bAp \qquad\quad 
\mbox{i.e. }\quad \bup_e=\frac e{mc^2} \bAp.             \label{bupExpl}
\ee
Eq. (\ref{bupExpl}), which holds also with the caret, allows to trade 
$\bup_e$  for $\bAp$ as an unknown. 
From (\ref{pump})  it follows  for $t\le 0$
\vskip-.5cm
\be
\bAp(t,z)=\Bap(ct\!-\!z),\qquad \Bap(\xi)\!\equiv\! -\!\!\int^{\xi}_{ -\infty }\!\!\!d\eta\:\Bep(\eta).
 \label{Apin}
\ee
\vskip-.2cm
The  local conservation \ $n_e\,dz=\widetilde{n_0}\,dZ$ \ of the number of electrons
(whence the continuity equation
) becomes 
\be
 n_e(t,z)=\widetilde{n_0}\!\left[Z_e(t,\!z)\right] \,\partial_z  Z_e(t,z),
\label{n_h}
\ee
and the Maxwell equations \ $\nabla\!\!\cdot\!\bE\!-\!4\pi j^0\!=\!\partial_z E^z\!-4\pi e(n_p-n_e)\!=\!0$, \
$ \partial_tE^z/c\!\!+\!4\pi j^z \!\!=\!( \nabla\!\!\wedge\!\bB)^z\!\!=\!0$ ($\bj\!=\!-en_e\bb_e$  is 
the current density) with the initial conditions imply  \cite{Fio14JPA}
\be
E^z(t,\! z)\!=\!4\pi e \!\left\{
\widetilde{N}(z)\!-\! \widetilde{N}[Z_e (t,\! z)] \right\}\!, \:\:\:
\widetilde{N}(Z)\!\equiv\!\!\!\int^Z_0\!\!\!\!\!\! d\eta\,\widetilde{n_{0}}(\eta).
 \label{explE}
\ee
Relations (\ref{n_h}-\ref{explE}) allow to compute $n_e,E^z$  explicitly in terms of 
the assigned initial density $\widetilde{n_0}$
and of the still unknown  $ Z_e(t,z)$ (longitudinal motion); 
thereby  they further reduce the number of unknowns. The remaining ones are $\bA\!^{{\scriptscriptstyle\perp}},\bx_e$ and $u_e^z$, or - equivalently - $s_e$.

Using the Green function of $\frac 1 {c^2}\partial_t^2\!-\!\partial_z^2$ one finds that the Maxwell equation \ $\Box\bAp=4\pi\bjp$ (in the Landau gauges) \& (\ref{Apin}) amount to the integral equation \cite{Fio14JPA}
\be
\ba{ll}
\displaystyle\bA\!^{{\scriptscriptstyle\perp}}(t,z)-\Bap(ct\!-\!z)=&\!\!\! \displaystyle
-  \frac{2\pi e^2}{mc^2} \!\!\!\int \!\!\!   d \eta   d\zeta   \left(\!\frac{n_e\bAp}{\gamma_e}\!\right)\!\left(\eta,\zeta\right)\\[8pt]
&\displaystyle\times \,\theta(\eta)\, \theta\left(ct \!-\!\eta\!-\!|z\!-\!\zeta|\right).  
\ea       \label{inteq1}
\ee
Abbreviating \   $v\!\equiv\!\hat\bu^{{\scriptscriptstyle\perp}}_e{}^2\!=\![ e\hat\bA\!^{{\scriptscriptstyle\perp}}/{mc^2}]^2$, \
$\hat\Delta(\xi,Z)\!\equiv\!\hat z_e(\xi,Z)\!-\!Z$, \
the remaining eq.s   (\ref{EOM})  take the form $\hat\bx^{{\scriptscriptstyle\perp}}_e{}'\!=\!
\hat\bu ^{{\scriptscriptstyle\perp}}_e/\hat s_e$ and
\be
\hat\Delta'\!=\displaystyle\frac {1\!+\!v}{2\hat s_e^2}\!-\!\frac 12, \quad 
\hat s'_e(\xi,\!Z)\!=\!\frac{4\pi e^2}{mc^2}\!\left\{ 
\widetilde{N}\left[Z\!+\!\hat\Delta\right] \!-\! \widetilde{N}(Z) \right\}\!, \label{heq1} 
\ee
with  initial conditions
\ $\hat\Delta(-Z,\!Z)\!=\!0$, $\hat s(-Z,\!Z)\!=\! 1$.  
Eq.s (\ref{heq1}) prevent $\hat s_e$ to vanish anywhere, consistently with 
(\ref{defs0}): if  $\hat s_e\!\downarrow\! 0$ then  rhs(\ref{heq1}a)  blows up and forces
$\hat\Delta$, and in turn $\hat s_e \!-\!1$, to abruptly grow again positive. 
By causality
 $\bA\!^{{\scriptscriptstyle\perp}}(t,z)\!=\!0$ if $ct\!\le\! z$, 
hence $v,\hat\Delta,\hat s_e-1$ remain zero until $\xi\!=\!0$, 
and we can shift  the initial conditions to
\vskip-.3cm
\be
 \hat \Delta(0,\!Z)\!=\!0,  \qquad\qquad
 \hat s_e(0,\!Z)\!=\! 1.  \label{heq2}
\ee
\vskip-.1cm
\noindent
Moreover, as the right-hand side (rhs) of (\ref{inteq1}) is zero for $t\le 0$, 
 we can still  use (\ref{Apin}),
and by (\ref{bupExpl}) approximate \  $\hat\bu^{{\scriptscriptstyle\perp}}_e\!=\! e\Bap/{mc^2}$, within {\it short} time intervals $[0,t_s]$ (to be determined {\it a posteriori}); 
$\hat\bu ^{{\scriptscriptstyle\perp}}_e$ and the forcing term $v$ thus
become known functions of $\xi$ (only), and 
 (\ref{heq1})  a family parametrized by $Z$ of {\it decoupled ODEs}.
For every $Z$ (\ref{heq1}) have the form of Hamilton equations \ $q'=\partial \hat H/\partial p$, $p'=-\partial \hat H/\partial q$  of a 1-dim system: \  $\xi,\hat\Delta, -\hat s_e$  play the role of $t,q,p$, and the Hamiltonian is {\it rational} in $\hat s_e$ and reads
\bea
\ba{l}
\displaystyle\hat H( \hat \Delta, \hat s_e,\xi;Z)\equiv  \frac{\hat s_e^2+1\!+\!v(\xi)}{2\hat s_e}
+ \U( \hat \Delta;Z),  \\[8pt]
\U( \Delta;Z)\!\equiv\!\frac{4\pi e^2}{mc^2}\left[
\widetilde{{\cal N}}\!\left(Z \!+\!  \Delta\right) \!-\!\widetilde{{\cal N}}\!(Z)\!-\! \widetilde{N}\!(Z) \hat \Delta\right],  \\[8pt] 
\displaystyle\widetilde{{\cal N}}(Z)\equiv
\int^Z_0\!\!\!d\zeta\,\widetilde{N}(\zeta)\!=\!\int^{Z}_0\!\!\!d\zeta\,\widetilde{n_0}(\zeta)\, (Z\!-\!\zeta).
\ea                                \label{hamiltonian}
\eea
\vskip-.1cm
\noindent
Eq.s (\ref{heq1}-\ref{heq2}) can be solved numerically, or by quadrature where
$\Bep(\xi)\!=\! 0$. Finally \ $\hat\bx^{{\scriptscriptstyle\perp}}_e{}'\!=\!
\hat\bu ^{{\scriptscriptstyle\perp}}_e/\hat s_e$  is solved by
\vskip-.2cm
\be
\hat\bx^{\scriptscriptstyle \perp}_e(\xi,\bX)-\bXp=\!\int^\xi_0\!\!\! d\eta \,\frac{\hat\bu^{\scriptscriptstyle \perp}_e(\eta)}{\hat s(\eta,Z)}.         \label{hatsol'}
\ee
By derivation we obtain several useful relations, e.g.
\vskip-.2cm
\be
\frac{\partial Z_e}{\partial z}(t,z)
=\left.\frac{\hat\gamma_e}{\hat s_e\, \partial_Z\hat z_e}\right\vert_{(\xi,Z)=\big(ct\!-z,Z_e(t,z)\big)} .\label{invZtoz}
\ee
\vskip-.1cm
\noindent
Hence the maps \ $\hat \bx_e(\xi,\cdot)\!:\!\bX\!\mapsto\!  \bx$, \
  $\bx_e(t,\cdot)\!:\!\bX\!\mapsto\!  \bx$ \ are invertible,  and the hydrodynamic 
approach justified, as long as $ \partial_Z\hat z_e\!>\!0$. \ 
From (\ref{n_h}), (\ref{invZtoz})
\vskip-.2cm
\be
n_e(t,z)\!=\!\left.\frac{\hat\gamma_e\,\widetilde{n_{0}}}{\hat s_e\,\partial_Z\hat z_e}\right\vert_{(\xi,Z)=\big(ct\!-z,Z_e(t,z)\big)}.         \label{expln_e}
\ee
\vskip-.1cm

We analyze the  motions ruled by (\ref{heq1}-\ref{heq2}). 
$\widetilde{N}(Z)$ grows with $Z$, and so does the rhs(\ref{heq1}b) with $\hat\Delta$.
As soon as $v(\xi )$ \  becomes positive  for $\xi\!>\!  0$,  then so do also
$\hat\Delta$ and $\hat s_e\!-\!1$: all electrons reached by the pulse start to oscillate transversely and
drift forward (pushed by the ponderomotive force); the $Z\!=\!0$ electrons leave behind themselves
a  layer of ions $L_t$ of finite thickness $\zeta(t)\!=\!\Delta(t,0)\!=\!\hat\Delta[\tilde\xi(t,0),0]$ 
completely deprived of electrons. 
   $ \hat s_e$ keeps growing  as long as $\hat\Delta\!\ge\!0$, making the 
rhs(\ref{heq1}a)  vanish at the first $\bar\xi(Z)\!>\!0$ where $\hat s_e^2(\bar\xi,\!Z)\!=\!1\!+\!v(\bar\xi)$
and become negative for  $\xi\!>\!\bar\xi$. Hence $\hat\Delta(\xi,\!Z)$ reaches a positive maximum 
at $\xi\!=\!\bar\xi(Z)$ and then starts decreasing 
towards negative values (electrons are attracted back by ions in $L_t$).
By  (\ref{Lncond}), $\hat\Delta(l,\!Z)\!\ge\! 0$ for all $Z$:  the pulse is completely inside 
the bulk before any electron gets out of it, i.e. before 
$L_t$ is refilled. For $\xi\!>\!l$ the (conserved)
energy  $h(Z)\!=\!1\!+\!\!\int^l_0\!\!d\xi v'(\xi)/\hat s(\xi,Z)$ \cite{Fio17JPA}
determines as usual $P(\xi;Z)$ and its path  as the level curve $\hat H\!=\!h(Z)$,
i.e.
\be
\frac s2\!+\!\frac{\mu^2}{2s}\!=\! \bar\gamma(\Delta;Z)\!\equiv\! h(Z)\! -\!\U(\Delta;Z),
\quad\: \mu\!\equiv\!\sqrt{1\!+\!v(l)}.                            \label{H-level-cycle}
\ee
The center $C\!\equiv\!(\Delta,s)\!=\!(0,\mu)$,  is the only critical point;
for slowly modulated pulses $ v(l)\!\simeq\!0$ 
and $\mu\!\simeq\!1$.
 Solving (\ref{H-level-cycle}) with respect to $s$ one finds the two solutions
\be
\spm(\Delta;Z)\equiv \bar\gamma(\Delta;Z)\pm\sqrt{ \bar\gamma^2(\Delta;Z)-\mu^2}; 
   \label{spm}
\ee
they fulfill $\sp\sm\!=\!\mu^2$. 
 \ The solutions $\Delta_{{\scriptscriptstyle M}}(Z)\!>\!0,\Delta_m(Z)\!<\!0$
of the equation  $h\!=\!\U(\Delta;Z)$ are the  maximal, minimal displacements. 
The maximum $s_{{\scriptscriptstyle M}}$, minimum $s_m$  of  $s$ are
\vskip-.2cm
\be
 s_{{\scriptscriptstyle M},m}(Z)
=\spm(0,Z)=h\pm\sqrt{h^2\!-\!\mu^2}. \label{smM}
\ee
\noindent
From (\ref{explE})  it follows $\hat E^z{}'\!\propto\! \hat \Delta'\!=\!0$
only when $\hat \Delta\!=\!\Delta_{{\scriptscriptstyle M}},\Delta_m$; hence the
maximum $ E^z_{{\scriptscriptstyle M}}$ and minimum $ E^z_{{\scriptscriptstyle M}}$ of
$E^z$ experienced by the $Z$-electrons are respectively given by
\be
E^z_{{\scriptscriptstyle M},m}(Z)\!=\!4\pi e \!\left\{
\widetilde{N}\left( Z\!+\!\Delta_{{\scriptscriptstyle M},m}\right)\!-\! \widetilde{N}(Z) \right\}.
 \label{E^zmM}
\ee
Since $\U(\Delta;0)\!=\!0$ for $\Delta\!\le\!0$, then $\hat\Delta(\xi,0)\!\to\!-\!\infty$ as $\xi\!\to\!\infty$: 
the $Z\!=\!0$ electrons escape to $z_e\!=\!-\infty$ infinity.  Whereas if $Z\!>\!0$ then $\U(\Delta;\!Z)\!\to\!\infty$ as 
$|\Delta|\!\to\!\infty$, the path is a cycle around $C$, and $P(\xi;Z)$ is periodic in $\xi$, with period
\bea
 c\, t_{{\scriptscriptstyle H}}=\xi_{{\scriptscriptstyle H}}=2\int
^{\Delta_{{\scriptscriptstyle M}}}_{\Delta_m}\!\!\!d\Delta\, \frac{\bar\gamma(\Delta;Z)}{\sqrt{ \bar\gamma^2(\Delta;Z)-\mu^2}}:                         \label{period}
\eea
all $Z\!>\!0$ electron layers do longitudinal oscillations about their initial positions. 
There are $Z_b\!>\!0$ and $Z_d\!>\!Z_b,z_s$ such that: i) The $Z\!<\!Z_b$ 
electrons exit and re-enter the bulk, while the $Z\!\ge\!Z_b$ electrons remain inside the bulk; 
their oscillations  arrange in a PW 
trailing the pulse.
ii) If $Z\!\ge\!Z_d$ 
then  for all $\xi$ $\hat z_e\!\ge\! z_s$, $\widetilde{n_{0}}(\!\hat z_e\!)\!\equiv\!n_0$,  $\U(\Delta,Z)\!\equiv\!M\Delta^2\!/2$, (\ref{heq1}) no longer depends on $Z$ and reduces to the  equation \cite{AkhPol56}
of a single forced, relativistic harmonic oscillator (formulated in an unusual way):
\be
 \Delta'=\displaystyle\frac {1\!+\!v}{2s^2}\!-\!\frac 12,\qquad
 s'=M\Delta, \quad M\!\equiv\!\ba{l}\!\! \frac{4\pi n_0e^2}{mc^2} \!=\!\frac{\omega_p^2}{c^2}\!\!\ea.\label{e1}
\ee
To illustrate, in fig. \ref{graphsb} we plot the solution induced by a  gaussian modulated 
$\Bep$ (parameters are chosen as in \cite{BraEtAl08}).
By the $Z$-independence of $\Delta,s,h$, \  $z_e(t,\!Z)$ has the inverse
\be
 Z_e(t,z)=z-\Delta(ct\!-\!z),                                                \label{sol"}
\ee 
making all Eulerian fields completely explicit and  dependent on $t,z$ only through $ct\!-\!z$, 
i.e. propagating as traveling-waves; in particular (\ref{explE}), (\ref{expln_e}) take the form
\bea
E^z(t,z)=4\pi e n_0 \, \Delta(ct\!-\!z), \qquad
\qquad\qquad\qquad\label{explE'} \\[6pt]
n_e(t,z)
=\frac{n_0}2  
\left[1\!+\! \frac{1\!+\!v(ct\!-\!z)}{s^2(ct\!-\!z)}\right]=
\frac{n_0}{1\!-\!\beta^z(ct\!-\!z)},      \label{expln_e'}
\eea
(as predicted  in \cite{AkhPol56}, formula (9) with phase velocity $V\!=\!c$),
implying \ $ n_e\!>\!n_0/2$, \ $n_e(t,z) \!\simeq\!n_0/2$ if $s^2(ct\!-\!z)\!\gg\! 1\!+\!v(ct\!-\!z)$.                 
Moreover \ $\Delta_{{\scriptscriptstyle M}}\!=\!-\Delta_m$ and 
 $h\!=\!M \Delta_{{\scriptscriptstyle M}}^2/2$; \ 
(\ref{E^zmM}) gives \
$ E^z_{{\scriptscriptstyle M}}\!=\!4\pi e n_0\Delta_{{\scriptscriptstyle M}}\!=\!- E^z_m$; \
 by (\ref{expln_e'}),  
the maximum $n_{{\scriptscriptstyle M}}$ of $n_e$ is obtained at $s=s_m$, as computed in  (\ref{smM}):
\bea
n_{{\scriptscriptstyle M}}=\frac{n_0}2 \left[1\!+\! \frac{\mu^2}{s_m^2}\right]
=\frac{n_0\,h}{\mu^2}\left[h\!+\!\sqrt{h^2\!-\!\mu^2} \right].
 \label{nM}
\eea
By (\ref{sol"}), if $Z\!>\!Z_d$   the map
$ z_e(t,\cdot):Z\!\mapsto\! z$ is invertible for all $t$,  thus
justifying the hydrodynamical picture used so far.
Collisions  can occur only between two electron layers with $Z\!<\!Z_d$. 
In \cite{Fio17} we will show that indeed the time $t_c(z_s)$ of the first collisions involves  $Z_c$-electrons 
(with some $Z_c\!<\! Z_d$) and  is earliest if $z_s=0$, i.e. $\widetilde{n_0}(Z)\!=\!n_0\theta(Z)\!=\!n_b\theta(Z)$
(a few corresponding $H\!=$const curves are plot  in fig. \ref{graphsb}d);
then $t_c(z_s)\!\ge\! t_c(0)> \frac 54 T\!\!_{{\scriptscriptstyle H}}\!+\! \frac {Z_c}c$. 
The collisions (leading to local  wave-breaking and dissipation of ordered kinetic energy into disordered one)
may be useful to inject part of the electrons in the hollows of the PW for acceleration purposes.
As known  \cite{BulEtAl98,BraEtAl08,LiEtAl13}, they may occur not only near where $\widetilde{n_0}(Z)$ grows, but also near where it
decreases. As an illustration, in fig. \ref{Worldlinescrossings} we plot the electron worldlines (WL)  for 
$0\!<\!ct\!<\!300\lambda$ under 
conditions as in section III.B of Ref. \cite{BraEtAl08}: $\widetilde{n_0}(Z)$ grows linearly from $0$ to 
$n_b\!=\!n_{cr}/250$ in $0\!<\!Z\!<\!120\lambda$, equals $n_b$  in $120\lambda\!\le\!Z\!\le\!190\lambda$,  decreases linearly from $n_b$ to $n_0=n_{cr}/400$ in $190\lambda\!<\!Z\!<\!200\lambda$, equals $n_0$
 for $Z\!\ge\!200\lambda$; the pump is linearly polarized,  gaussian-modulated with
normalized peak  amplitude $a_0\!\equiv\!\lambda eE^{\scriptscriptstyle \perp}_{\scriptscriptstyle M}/mc^2\!=\!2$
and full width at half maximum intensity $l_{fwhm}\!=\!10.5\lambda$. The drift of the small-$Z$ electrons
is at the base of the slingshot effect (see next section). Collisions 
involve electrons both in the up- and down-ramp. The latter
are more gentle, i.e.  WL intersect with very small angles; in \cite{BraEtAl08} it was shown that
the resulting self-injection of electrons in the PW is 'optimal' for their WFA. If the down-ramp were longer, collisions there would occur after more oscillations.

Summarizing,  for \ $t\!\le\! t_c(z_s)$ \ no collisions occur,  the maps   $z_e(t,\cdot)\!: Z\!\mapsto\! z$
are invertible, and the hydrodynamic description is justified.
For $t\!>\! t_c$ collisions can occur only near the density inhomogeneities;
the associated perturbations cannot reach the part of the PW just behind the pulse, as this travels
with phase velocity $c$.
On the other side, the $Z\!\simeq\! 0$ electrons  go far backwards before coming back, so
are also not affected for  long $t$.

Finally,  approximating   $\bAp(t,z)\!\simeq\! \Bap(ct\!-\!z)$   is acceptable as long as the so determined motion makes \ $|\mbox{rhs}(\ref{inteq1})| \!\ll\! |\Bap|$; in the region of interest here this is the case. Otherwise replacing $\bAp\mapsto \Bap(\xi)$ into rhs(\ref{inteq1}) determines the first correction to $\bAp$; replacing the corrected $\bAp$ into (\ref{bupExpl}) and the new $\hat\bup_e$ into  (\ref{heq1}-\ref{heq2})  one can determine 
the motion  with more accuracy; and so on.

\section{Finite $R$ and discussion}
\label{3D-effects}

\begin{figure}
\includegraphics[width=6cm]{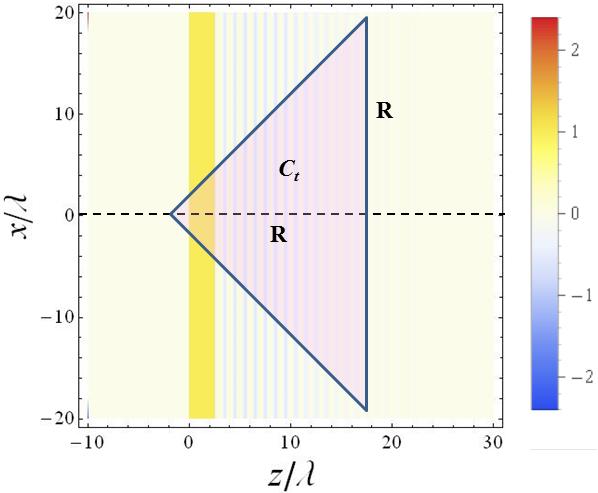}
\caption{Normalized total charge density $1\!-\!n_e/n_0$ at $ct=19\lambda$ (conditions as in 
fig. \ref{graphsb}).
The  electron density `hole' includes the intersection of the  ion layer $L_t$ (yellow) and  causal cone $C_t$ (pink) behind the pulse.}
\label{CCone}       
\end{figure}

By causality, if two solutions of the dynamic equations coincide in a 
spacetime region $\D$, then they coincide also in the {\it future Cauchy development} $D^+(\D)$
(the set of all points $x$ for which every past-directed causal line through $x$  intersects $\D$).
Hence, knowing one solution determines also the other  
within $D^+(\D)$.  
Here all the dynamical variables are exactly known at $t\!=\!0$, and also for $0\le\!ct\!\le\!z$
(there the plasma is still at rest and $\bE,\bB$  zero).
We adopt: i) as $\D$ the surface $\D^0_{{\scriptscriptstyle R}}$
 of equations $\rho\!<\! R$ and either $\!t\!=\! 0$, $z\!<\!0$, or $0\le\!ct\!\le\!z$; ii)
as the known solution the plane one of section \ref{Planewavessetup};
iii) as the unknown solution  the ``real'' one induced by 
(\ref{pump}). Hence, 
 at all $t$ all dynamical variables, in particular $n_e$ (see fig. \ref{CCone}), are strictly the same 
as in section \ref{Planewavessetup}, within the causal cone \
$\C_t\!\equiv\!\{(\rho'\!,\varphi'\!,z') \:\:|\:\: 0\!\le\!\rho'\!+\!ct\!-\!z'\!\le\!R\}$ \ (in cylindrical coordinates)
trailing the  pulse, and approximately the same, in a neighbourhood of it. 
Hence, for small $t$ there is a  ``hole'' $h_t$ in the electron
distribution including at least $\C_t\cap L_t$, while for larger $t$ the PW behind the pulse 
is the same  inside $\C_t$.
If $R$ is `large', wave-breaking around the vacuum-bulk interface takes place also
within $\C_t$. Whereas for smaller $R$ fulfilling
\be
R\!\gg\!\Delta\bxp_e, \quad t_e \!-\!\bar t \sim \frac Rc,   \quad
r \equiv R-\frac{\zeta(t_e\!-\!l/c)}{2(t_e\!-\!\bar t)} >0                                   \label{req}
\ee
($\bar t,t_e$ are the times of maximal penetration and expulsion  of the $Z\!=\!0$ electrons)
the $0\!\le\!Z\!\le\!Z_b$, $\rho\!\le\! r\!\le\! R$ electrons
exit  the bulk shortly after $C_t$ has completely entered it. 
Conditions (\ref{req}) respectively ensure:  that
these electrons move approximately as in section \ref{Planewavessetup} until their expulsion; 
that  they are expelled before
lateral electrons (LE), which are initially located outside the surface $\rho\!=\!R$ 
and are attracted towards the $\vec{z}$-axis, obstruct them the way out, colliding with each other.
The expelled electrons are decelerated by the electric force generated by
the net positive charge located at their right within $\rho\!<\!r$, which decreases as
$1/z_e^2$ as $z_e \!\to\! -\infty$; this allows the backward escape  
of a bunch of electrons with high energy ($1\!\div\! 5$MeV for a gaussian pulse of energy $\E\!=\!5$J, 
$l_{fwhm}\!\simeq\! 7.5 \mu$m,  $\lambda\!\simeq\! 0.8 \mu$m, spot size $R\!=\!4\!\div\!16\mu$m on a helium jet target), well collimated 
($u^{{\scriptscriptstyle \perp}}_f\!\simeq\!0$) if $\Bep$ is slowly modulated ({\it slingshot effect}). 
For more details see \cite{FioDeN16}.  

If $R$ is even smaller  the  LE  attracted towards the $z$-axis $\vec{z}$
collide closing part of $L_t$ into a (possibly temporary)  electron cavity (where $n_e\!=\!0$)  \cite{RosBreKat91,PukMey2002}
 before  any electrons are expelled backwards.  If  $R\!\lesssim\!\Delta\bxp_{e{\scriptscriptstyle M}}\!\!\equiv$the maximal transverse oscillations (\ref{hatsol'}),  the solution of section \ref{Planewavessetup} is unreliable even for the $\bX\!=\!(0,0,Z)$ electrons.

We can make the results more explicit  if $\Bep$ in (\ref{pump}) is a monochromatic wave 
modulated by some $\epsilon(\xi)\ge 0$,
\be
\Be^{{\scriptscriptstyle \perp}}\!=\!\epsilon 
\Be_o^{{\scriptscriptstyle \perp}}, \quad
\Be_o^{{\scriptscriptstyle \perp}}\!(\xi)\!=\!\bi a_1\cos (k\xi\!+\!\varphi_1)+\bj a_2\sin (k\xi\!+\!\varphi_2),
 \label{modulate}
\ee
with support $[0,l]$ ($a_1^2\!+\!a_2^2\!=\!1$).
If $f(\xi)$ is a regular function  vanishing $\xi\!=\!-\infty$ integration by parts gives
\be
\int^\xi_{-\infty}\!\!\!\!\!\! d\eta\: f(\eta)e^{ik\eta} = -\frac ik f(\xi)e^{ik\xi}+O\!\left(\frac 1{k^2}\right); \label{oscillestimate}
\ee
(the remainder $O(1/k^2)$ is `small' if $|f'|\!\ll\! |k f|$, see Appendix 5.4  of \cite{Fio17}).
If $\Bep$ is slowly modulated (i.e. $|\epsilon'|\!\ll\! |k\epsilon|$ on $[0,l]$) then
\ $\Bap(\xi)\!\simeq\!  \epsilon(\xi) \,\Bep_o(\xi\!+\!\pi/2k)$; 
\ hence $\Bap(\xi),\bup(\xi) \!\simeq\!   0$ if $\xi\!>\!l$. Since $|f'|\!\ll\! |k f|$ holds also 
for $f\!=\!\hat s_e,\partial\hat z_e/\partial Z$, (\ref{hatsol'}) yields \ $\Delta\bxp_e\!\simeq\!-e\Bep/k^2mc^2\hat s_e$, and using (\ref{expln_e}) one can easily estimate  $\mbox{rhs}(\ref{inteq1})$, so as to check  
the condition $R\lesssim |\Delta\bxp_{e{\scriptscriptstyle M}}|$
and the approximation $\bAp\!\simeq\!\Bap$.


\end{document}